\documentclass{mem}
\usepackage{natbib}\usepackage{txfonts}\usepackage{balance}
\usepackage{graphicx}
\usepackage[a4paper,breaklinks,dvipdfm]{hyperref}
\idline{75}{282}
\begin{document}
\def\teff{$T\rm_{eff }$}
\def\kms{$\mathrm {km s}^{-1}$}

\title{
Irradiated brown dwarfs
}

   \subtitle{}

\author{
S.L. \,Casewell\inst{1}, M.R. Burleigh\inst{1}, K.A. Lawrie\inst{1}, P.F.L. Maxted\inst{2}, P.D. Dobbie\inst{3}, R. Napiwotzki\inst{4}
          }

  \offprints{S. L. Casewell \email{slc25@le.ac.uk}}

\institute{
Department of Physics and Astronomy,
University of Leicester, University Road, 
Leicester, LE1 7RH, UK
\and
Astrophysics Group, Keele University, Staffordshire, ST5 5BG, UK.
\and
School of Mathematics and Physics, University of Tasmania, Hobart, Tasmania 7001, Australia.
\and
Science \& Technology Research Institute, University of Hertfordshire, College Lane, Hatfield, AL10 9AB, UK.
}

\authorrunning{Casewell et al}

\titlerunning{Irradiated brown dwarfs}

\abstract{
We have observed the post common envelope binary WD0137-349 in the near infrared $J$, $H$ and $K$ bands and have determined that the photometry varies on the 
system period (116 min). The amplitude of the variability increases with increasing wavelength, indicating that the brown dwarf in the system is likely being irradiated by its 16500 K white dwarf companion.  The effect of the (primarily) UV irradiation on the brown dwarf atmosphere is unknown, but it is possible that stratospheric hazes are formed.  It is also possible that the brown dwarf (an L-T transition object)  itself is variable due to patchy cloud cover. Both these scenarios are discussed, and suggestions for further study are made.

}
\maketitle{}

\section{Introduction}

There has long been a known dearth of brown dwarf companions to solar-type stars with orbital periods $<$5 years (equivalent to orbital separations $<$3AU) when compared with lower mass planetary companions or more massive stellar companions \citep{grether06}.  For instance   the WASP and Corot surveys observed tens of thousands of bright stars (V$\sim$15) and only discovered 2 transiting brown dwarfs: WASP-30b \citep{anderson} and Corot-15B: \citep{bouchy} compared to over 50 planets. This scarcity of objects is known as the brown dwarf desert, and may in fact, extend to much larger separations \citep{metchev}.  The reason for the lack of brown dwarf companions at these separations is unknown, but it is likely related to the formation mechanisms involved.

In these systems the main sequence star is bright and there is low contrast between the star and the brown dwarf. An alternative is to search for these binaries in their highly evolved form: white dwarf - brown dwarf binaries.    However, detached brown dwarf and very
low-mass stellar companions
to white dwarfs are rare; the fraction of L-type secondaries to white dwarfs
is just 0.4$\pm$0.3\% \citep{steele11}. Proper motion surveys and searches for
IR excesses have so far found only a handful of confirmed examples
(\citealt{becklin88, farihi04, maxted06, steele07, steele09, burleigh11,
  dayjones11, debes11, casewell12, steele13}), none of which have a reliably determined age
independent of the white dwarf parameters.  Only in 4 systems,  WD0137-349B
(L8, 0.053M$_{\odot}$, P$_{\rm orb}$=116 min; \citealt{maxted06}), NLTT5306 (L4-L7, 0.056M$_{\odot}$, P$_{\rm orb}$=101.88 min; \citealt{steele13}) , WD0837+185 ($>$T8, $\sim$30M$_{\odot}$, P$_{\rm orb}$=4.2 hr; \citealt{casewell12}) and GD1400B (L6-L7,
0.07-0.08M$_{\odot}$;P$_{\rm orb}$=9.98hrs \citealt{farihi04,burleigh11})  is the brown dwarf known to have survived a phase of common envelope  evolution. This phase of binary star evolution involves the brown dwarf being engulfed by, and immersed in, the expanding atmosphere of the white dwarf progenitor as it evolves away from the main sequence \citep[see e.g.][]{davis12}. 

These systems are very close, and likely tidally locked. In two systems,  WD0837+185 and in NLTT5306 we see evidence of sinusoidal modulations on the orbital period in the optical $V$ band for WD0837+185 \citep{casewell12} and in the $I$ band for NLTT5306  \citep{steele13}.

\section{Observations}

To further investigate the effects of irradiation on the brown dwarf WD0137-349B, we obtained a near-infrared K-band lightcurve with IRIS2 on the AAT in service time July 2006, covering the entire ~2 hour orbital period. Unfortunately, the observing conditions were poor, with seeing between 3-4 arcseconds. None-the-less, the data appeared to show variability at a level of over +/-10\%. This degree of modulation was unexpected.

We repeated the observations  on the night of 19 July 2007 and  data were obtained with IRIS2 at $J$, $H$ and $K_s$ across two successive binary orbits (ie for a total of four hours) by cycling repeatedly through the three filters with 10s exposures and a 5 point jitter pattern. In this manner, a data point was obtained in each waveband every 2-3 minutes.  The seeing was 2" and the S/N per image was $\sim$40.
The data were reduced using the \textsc{starlink orac-dr} IRIS2 pipeline which dark corrected, flat fielded and mosaiced the jittered images. Photometry was performed  using the \textsc{starlink autophotom} routine using an 8 pixel aperture and differential magnitudes were measured with respect to 3 non-variable stars of similar brightness in the field of view. We folded the data in each waveband onto the ephemeris which was derived from 70 radial velocity measurements made over 2 years with UVES on the VLT (Casewell et al., in prep). The radial velocity ephemeris leads the data by 0.25 of a phase as is expected.

\begin{figure*}
\begin{center}
\scalebox{0.4}{\includegraphics[angle=270]{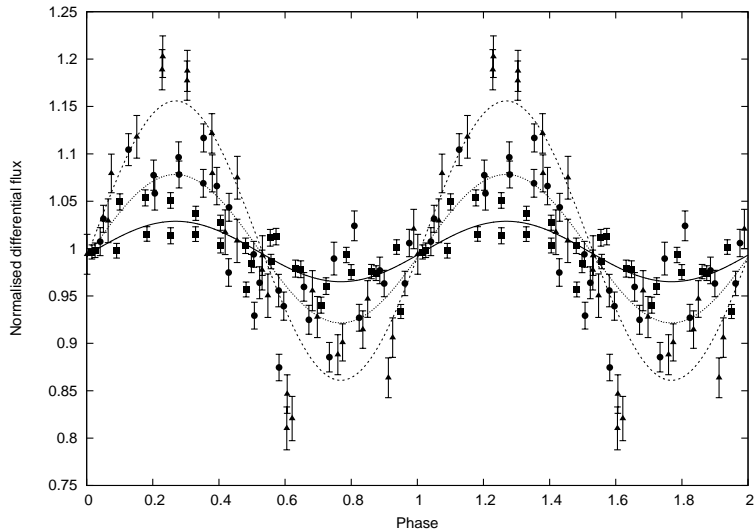}}
\caption{
\footnotesize
$J$ (boxes; solid line), $H$(circles; dotted line) and $K_s$(triangles; dashed line) lightcurves of WD0137-349B observed with IRIS2. These data have been folded onto the ephemeris of the system, offset by 0.25 of a phase. The best fit sine curves are shown by the solid, dotted and dashed lines.
}
\label{lightcurves}
\end{center}
\end{figure*}

The resultant light curves are shown in Figure \ref{lightcurves}.
\section{Discussion}

\begin{figure*}
\begin{center}
\scalebox{0.4}{\includegraphics[angle=270]{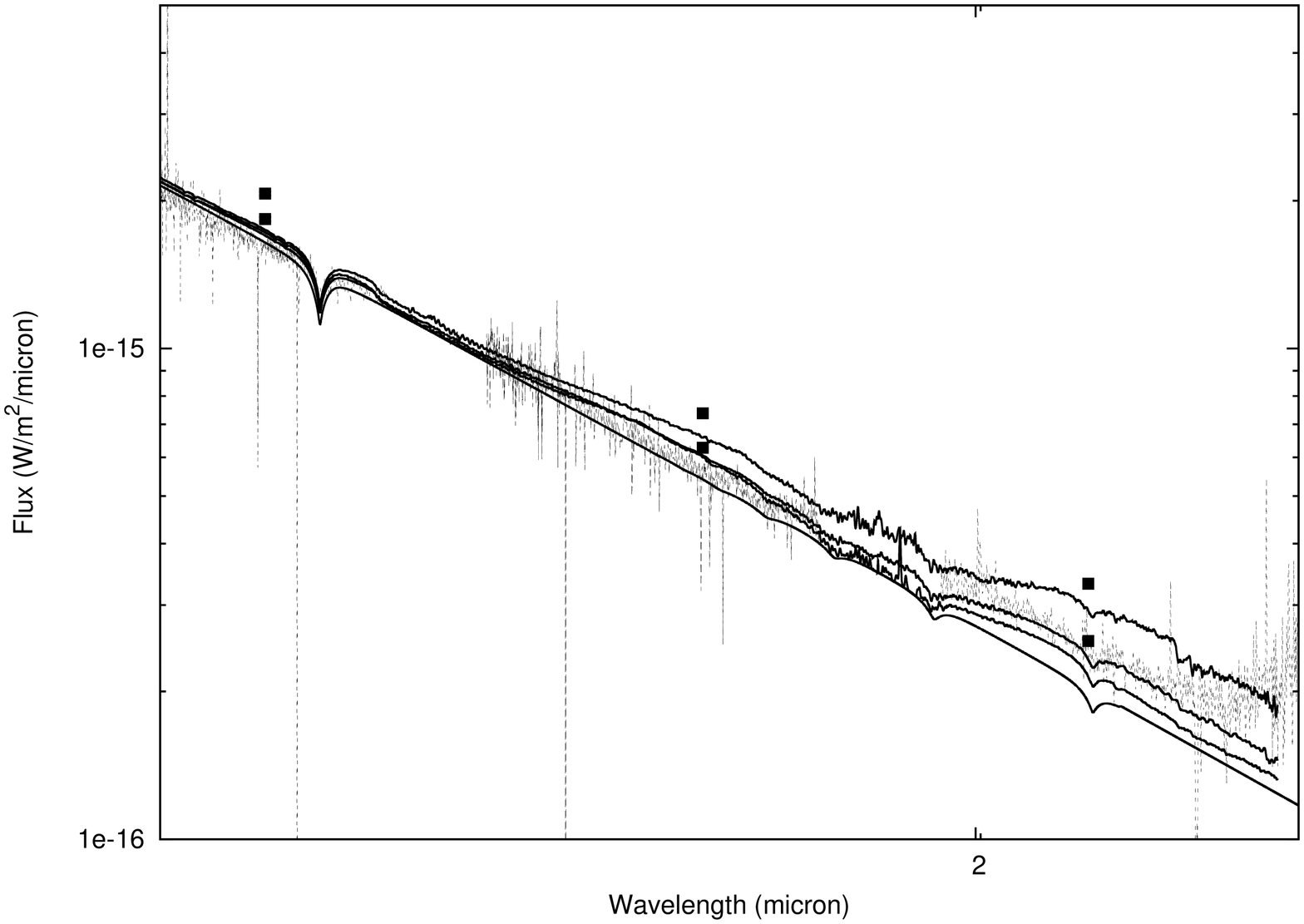}}
\caption{
\footnotesize
$J$, $H$ and $K_s$ magnitudes at maximum and minimum brightness superimposed onto the GNIRS spectrum of WD0137-349AB (grey line)  presented in \citep{burleigh06} as well as a synthetic white dwarf spectrum of WD0137-349 combined with a L6, and L8 and a T2 dwarf (black lines fro highest to lowest flux). The bottom line is the synthetic white dwarf spectrum
}
\label{spectra}
\end{center}
\end{figure*}

These data show variability which is in phase in all three near-infrared filters across the binary orbit. The peak-to peak values are: 6.4$\pm$0.4\% at $J$,  15.7$\pm$0.9\% at $H$, and increasing to 29.5$\pm$1.2\% at $K_s$. There are clearly differences between the day and night hemispheres of the brown dwarf, causing this variability. As with hot Jupiters, we expect the brown dwarf to be synchronously rotating and perpetually displaying one hemisphere to the 16,500K white dwarf. In \citet{burleigh06} we made a crude estimate of the temperature difference between the two hemispheres by comparing the maximum and minimum fluxes. In Figure \ref{spectra}, we have plotted these alongside models for the white dwarf + substellar companions from a spectral type of L0 to T2. At minimum in the $K_s$ band, the flux is consistent with an L8 brown dwarf. At maximum, the data favour an earlier, L6 classification. The difference in temperature between these two spectral types is 200-300K \citep{golimowski04}. Therefore, the dayside seems to be heated to a temperature around 1600-1700K, and the nightside is a cooler 1300-1400K. 

\citet{burleigh06} determined that the GNIRS spectra of WD0137-349 is best matched with a white dwarf + L8 brown dwarf composite model, suggesting that it was obtained when the night hemisphere was facing us.  The ephemeris derived from the radial velocity measurements of the white dwarf also confirm this. We note that for ages $>$1Gyr , an L8 spectral type and a temperature of 1300-1400K is entirely consistent with the measured mass of  the brown dwarf, as predicted by evolutionary models.  

As the variability increases as we progress from $J$ through to $K_s$, this suggests that in each waveband we are probing different heights or chemical species in the atmosphere. Cloud particles increase in size as we move deeper within the cloudy atmosphere, and photons are scattered from the cloud at greater depths with longer wavelength\citep{marley99}.

\citet{marley99} suggest that planets and brown dwarfs should be most reflective in the optical, and dark in the near-IR, if their atmospheres are clear. For cloudy objects, however, they are more reflective in the IR possibly due to hazes in the upper atmosphere. They also note that incident UV radiation (as in the case where a brown dwarf is being irradiated by a white dwarf) will drive photochemical reactions that can produce stratospheric hazes. These hazes,  however, have the greatest effect at T$_{\rm eff}<$1000K \citep{zahnle10}, cooler than WD0137-349B.

While it is likely that the variability is being caused by the effects of irradiation, we cannot discount the fact that some brown dwarfs on the L-T transition are variable due to patchy cloud cover \citep{radigan12, burgasser02, ackerman01}. Indeed the object 2MASS21392676+0220226 shows 60\% peak-to-peak variability in the $J$ band, and variability that ranges between 17\% and 60\% in the $J$ and $K$ band. In this case, it seems that the variability is, indeed, variable.  The observed  variability of WD0137-349B is in phase across the $J$, $H$ and $K_s$ bands which suggests that a patchy atmosphere is unlikely, although more data obtained over multiple orbits is required before we can determine if the variability evolves.

\section{Conclusions}
We have shown that the brown dwarf WD0137-349B is being irradiated by it's 16 500 K white dwarf companion which is seen as variability in the near-IR that increases in amplitude with wavelength. As the two components are likely tidally locked, this irradiation is resulting in a hotter "dayside" and cooler "nightside" on the brown dwarf.  The temperature difference between the two hemispheres is 200-300K, and the irradiated side, may show some sort of photochemical haze.

Obtaining lightcurves for entire orbits in both the near- and mid-IR would greatly increase our understanding of the system.  In the mid-IR particularly, on Jupiter regions of low condensate opacity are seen as bright hot spots at 5$\mu$m. \citet{gelino00} predicted that this mechanism could lead to variability as high as 20\%.  It would also be useful to obtain data over multiple orbits to determine if the variability is variable, which would indicate a non-homogenous cloud structure.

\begin{acknowledgements}
SLC acknowledges  support from the College of Science and Engineering at the University of Leicester.
This work made use of data from the Anglo-Australian Telescope. 

\end{acknowledgements}

\bibliographystyle{aa}

\end{document}